\documentclass[usenatbib]{mn2e}
\usepackage{psfig}

\newcommand{\spose}[1]{\hbox to 0pt{#1\hss}}
\newcommand{\approxpropto}{\mathrel{\spose{\lower 3pt\hbox{$\sim$}}
	\raise 2.0pt\hbox{$\propto$}}}
\def\approxgt{\mathrel{\spose{\lower 3pt\hbox{$\sim$}}
	\raise 2.0pt\hbox{$>$}}}
\def\approxlt{\mathrel{\spose{\lower 3pt\hbox{$\sim$}}
	\raise 2.0pt\hbox{$<$}}}

\newcommand{\dd}{\mbox{\rm d}}
\newcommand{\etal}{et al.{}}
\newcommand{\keV}{\mbox{\rm keV}}
\newcommand{\MDel}{{\mbox{$M_\Delta$}}}
\newcommand{\Mdel}{{\mbox{$M_{2500}$}}}

\newcommand{\Msun}{\mbox{$\rm M_{\odot}$}}
\newcommand{\mumh}{\mbox{$\mu m_{\rm H}$}}
\newcommand{\Nbody}{\mbox{$N$-body}}
\newcommand{\calPDel}{{\mbox{${\cal P}_\Delta$}}}
\newcommand{\Pgas}{\mbox{$P_{\rm gas}$}}
\newcommand{\rDel}{{\mbox{$r_\Delta$}}}
\newcommand{\rdel}{{\mbox{$r_{2500}$}}}

\newcommand{\rhogas}{\mbox{$\rho_{\rm gas}$}}
\newcommand{\Tgas}{\mbox{$T_{\rm gas}$}}

\newcommand{\Zsun}{\mbox{$\rm Z_{\odot}$}}

\title[Temperature--mass relation for galaxy clusters]
{Can simulations reproduce the observed temperature--mass relation for
clusters of galaxies?}

\author[P.~A.~Thomas \etal]
{Peter A.~Thomas\thanks{E-mail: p.a.thomas@sussex.ac.uk},
Orrarujee~Muanwong, Scott T.~Kay and Andrew R.~Liddle \\
Astronomy Centre, University of Sussex, Falmer, Brighton, BN1\,9QJ
}

\date{December 19, 2001}

\begin{document}
%\journal{Preprint astro-ph/0112449} 
 
\maketitle

\begin{abstract}
It has become increasingly apparent that traditional hydrodynamical
simulations of galaxy clusters are unable to reproduce the observed
properties of galaxy clusters, in particular overpredicting the
mass corresponding to a given cluster temperature. Such overestimation
may lead to systematic errors in results using galaxy clusters as
cosmological probes, such as constraints on the density perturbation
normalization $\sigma_8$. In this paper we demonstrate that inclusion
of additional gas physics, namely radiative cooling and a possible
preheating of gas prior to cluster formation, is able to bring the
temperature--mass relation in the innermost parts of clusters into
good agreement with recent determinations by Allen, Schmidt \& Fabian
using {\it Chandra} data.
\end{abstract}


\section{Introduction}

Reproducing the observed number density of rich galaxy clusters has
long been thought to be one of the most reliable constraints on the
matter power spectrum on short scales. It has been studied by many
authors over the years (\citealt{E89}; \citealt{HA91};
\citealt*{WEF93}; \citealt*{EKF96}; \citealt{VL96}, \nocite{VL99}1999;
\citealt{H97}, \nocite{H00}2000; \citealt{BSB00}; \citealt*{PSW01};
\citealt{W01}), recent determinations typically yielding $\sigma_8
\sim 0.9$ to $1.0$ for the currently-favoured $\Lambda$CDM model with
matter density $\Omega_0 \simeq 0.3$.  However, recently evidence has
begun to accumulate from a number of sources that this may be a
significant overestimate, perhaps by tens of percent. For example, the
required $\sigma_8$ estimated from the 2dF galaxy survey
(\citealt{L02}; \citealt{V02}), or from that survey combined with
other probes (\citealt{E02}), is significantly lower, and there are
now several papers using galaxy clusters that also give lower results
(\citealt{BRT01}; \citealt{RB02}; \citealt*{VNL02}).

A low value was also found recently by \citet{S02}, who used an
observed relation between cluster temperature and mass \citep*{FRB01}
rather than one derived from hydrodynamical simulations.  This last
result is particularly significant, and points to the increasingly
evident result that traditional hydrodynamical simulations, which
include only adiabatic gas heating during collapse, are unable to
reproduce the observed properties of clusters. For example, the recent
{\it Chandra} results of \citet*[][hereafter ASF01]{ASF01} indicate
that, at least in the inner regions where data exists, clusters are
considerably hotter for a given mass than predicted by adiabatic
simulations.

Here we address the question of whether the inclusion of
additional gas physics, both radiative cooling of the gas and
preheating of the gas before cluster formation, is capable of bringing
the simulations into agreement with observations.  We concentrate only
on the inner regions of clusters, for which temperature profiles have
been measured by {\it Chandra}, and we find that indeed the
observations can be reproduced. In itself this is not sufficient aid to
theorists seeking to constrain $\sigma_8$, which requires an accurate
description of clusters out to the virial radius, but this encouraging
result suggests that simulations may soon be useful for this
purpose.  We will explore the cluster 
temperature--mass relation out to larger radii in a forthcoming paper.

\section{Determining cluster masses}

\subsection{Using the Hydrostatic Equation to measure masses}

The distribution of hot gas in a cluster can be used to measure its mass.
The intracluster medium is generally assumed to be in hydrostatic
equilibrium in a spherically-symmetric, static potential and the mass
is determined by balancing pressure support against the
gravitational attraction:
\begin{equation}
{1\over\rhogas}{\dd\Pgas\over\dd r}=-{GM(<r)\over r^2}\,,
\label{eq:hydrostatic}
\end{equation}
giving
\begin{equation}
M(<r) = {k\Tgas(r)\over\mumh G}\,
          \left|{\dd\ln\Pgas\over\dd\ln r}\right|\,r\,,
\label{eq:masshydro}
\end{equation}
where $\Pgas=(k/\mumh)\rhogas\Tgas$ is the pressure
of the intracluster medium, $\rhogas$ its density, $k$ is the
Boltzmann constant, $G$ the gravitational constant, and
$\mumh\simeq1.0\times10^{-24}$g the mean mass per particle.

Where \Pgas\ is simply the thermal pressure, then \Tgas\ corresponds
to the gas temperature.  One should really include all forms of
pressure support for the gas: kinetic (i.e.~turbulence), magnetic,
coupled relativistic particles, etc..  We have checked in our
simulations that the contribution from kinetic motions within the
inner regions of clusters is small and so for clarity of presentation
we stick with the thermal pressure in this paper.

Note that the determination of the mass within radius $r$ depends only
upon the properties of the gas at that radius; in particular the mass
determination is not affected by conditions in the outer parts of
clusters where the properties are poorly determined. We show in
Section~\ref{sec:hydro} that the use of equation~(\ref{eq:masshydro})
leads to good estimates of the mass within regions accessible to {\it
Chandra} observations.  This lends credence to the mass determinations
from X-ray observations that we use in this paper.

\subsection{Observed and simulated temperature--mass relations}

If we apply equation~(\ref{eq:masshydro}) to measure the mass,
\MDel, within a radius, $\rDel$, for which the enclosed
density is $\Delta$ times the critical density, then
\begin{equation}
{kT(\rDel)\over\keV}\approx 8.3\,
\left(\Delta\over200\right)^{1\over3}\left(\calPDel\over2\right)^{-1}
\left(\MDel\over10^{15}h^{-1}\Msun\right)^{2\over3}\,,
\label{eq:tm}
\end{equation}
where $\calPDel=-\dd\ln\Pgas/\dd\ln r$ measured at \rDel.

For self-similar clusters, for which \calPDel\ is independent
of mass and $T(\rDel)$ is a constant multiple of the observed
temperature, $T_X$, we recover a scaling relation 
$T_X\propto\MDel^{\!\!2/3}$.
In order to compare different results we will write
\begin{equation}
{kT_X\over\keV}=A_\Delta
\left(\MDel\over10^{15}h^{-1}\Msun\right)^{\alpha}\,.
\label{eq:tmnorm}
\end{equation}
$A_\Delta$ gives the relative normalization of the relations; if
$\alpha\neq{2/3}$ then this comparison will be exact only for
$\MDel=10^{15}h^{-1}\Msun$.  Theorists can most easily predict the
mass within the virial radius of cluster, corresponding to
$\Delta\simeq111$ for our chosen cosmology, but X-ray observations do
not extend to such large radii and so some degree of extrapolation
(i.e.~the extension of a mass-model beyond the range of the
observational data) is usually required.

A useful summary of simulated and observational results is given
in \citet{AC02}.  The main simulated catalogues are by \citet{EMN96},
\citet{T01} and \citet{BN98}.  The former two, using smoothed-particle
hydrodynamics (SPH), found $A_{200}\approx8.0$, whereas the latter, using an
Eulerian grid-code found $A_{200}\approx7.0$.  In a more recent paper,
in which they considered a variety of definitions of X-ray temperature
for an ensemble of 24 highly-resolved SPH clusters, \citet{ME01} agreed
with this lower normalization.  None of these simulations included
radiative cooling.

By contrast the observational results using resolved
surface-brightness and temperature profiles from {\it ROSAT} and {\it
ASCA} have higher normalizations.  \citet*{XJW01} find
$A_{200}\approx9.4$, whereas \citet*{HMS99} and \citet{FRB01} both
find $A_{200}\approx11.2$.

Despite the uncertainty in both the simulated and the observational
results, it is clear that the observed normalization of the
temperature--mass relation significantly exceeds the simulated one.  

\subsection{Observational determination of \Mdel}

In this paper, we attempt a partial resolution of the differences
between simulations and observations discussed in the previous
section.  In doing this, we concentrate on the results of ASF01.
The reasons for this are fourfold:
\begin{enumerate}
\item {\it Chandra} observations give the best available
estimates of density and temperature profiles for the X-ray emitting
gas.
\item The mass estimates are, mostly, backed up by observations and
modelling of gravitationally-lensed background galaxies (arcs).
\item They present results for the mass-weighted temperature of the
gas.  The use of mass-weighted rather than emission-weighted
temperatures greatly simplifies the comparison of simulations and
observations.
\item They do not attempt to extrapolate their results beyond the
radius that is accessible to observations.
\end{enumerate}
ASF01 measured the mass and temperature of 5 clusters within
\rdel.  They found a best-fit slope for the mass-temperature relation
that is consistent with the self-similar value of 1.5.  We
rewrite their relation here with mass as the ordinate as we are
complete in mass rather than temperature in our simulations:
\begin{equation}
{kT_{2500}\over\keV}\approx
   19.2\,\left(\Mdel\over10^{15}h^{-1}\Msun\right)^{2\over3}\,,
\end{equation}
where $T_{2500}$ is the mass-weighted gas temperature within \rdel.

\section{Simulations}

The simulations that we discuss in this paper were carried out using
the Hydra \Nbody/hydrodynamics code (\citealt*{CTP95}; \citealt{PC97})
on the Cray T3E computer at the Edinburgh Parallel Computing Centre as
part of the Virgo Consortium programme of investigations into
structure formation in the Universe.  They will be described fully in
a longer paper (in preparation) and so we just summarize the
properties here.  Note that the simulations are very similar to those
discussed in an earlier paper, \citet{MTK01}, but the
parameters have been slightly adjusted so as to give a better fit to
the observed luminosity--temperature relation of clusters.

In this paper we present results for a single cosmology with density
parameter $\Omega_0=0.35$, cosmological constant $\Omega_{\Lambda
0}=0.65$, baryon density $\Omega_{{\rm b}0}=0.038$, Hubble parameter
$h=0.71$, cold dark matter density fluctuation parameter
$\Gamma=0.21$, normalization $\sigma_8=0.90$ and gravitational
softening $s=25\,h^{-1}$kpc.  160$^3$ particles each of gas and dark
matter were used in a box of side 100$h^{-1}$Mpc, giving particle
masses of $m_{\rm gas}\approx2.6\times10^9h^{-1}\Msun$ and $m_{\rm
dark}\approx2.1\times10^{10}h^{-1}\Msun$, respectively.

\begin{figure}
\psfig{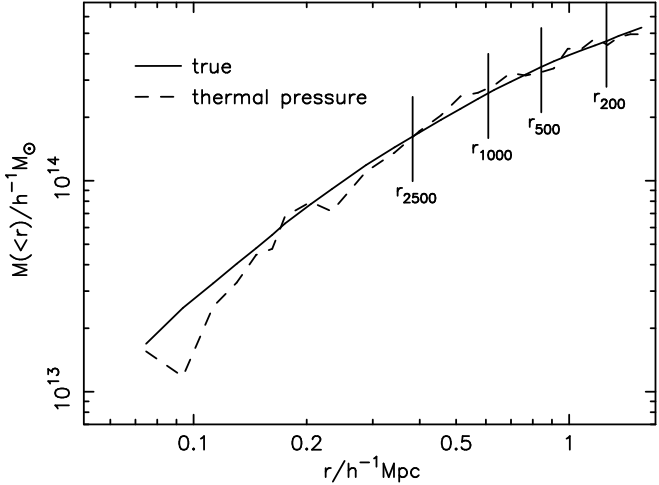}
\caption{The solid line shows the actual mass distribution in the
   cluster; the dashed line shows the mass from
   equation~(\ref{eq:masshydro}) using the thermal pressure.}
\label{fig:hydro}
\end{figure}

Three simulations were undertaken, each with identical initial
conditions so that the clusters that they produce are directly
comparable, but with different cooling properties:
\begin{description}
\item[\bf Non-radiative:] An adiabatic run included for testing and 
comparison
purposes only.  As discussed in \citet{MTK01}, the gas in the
centres of the clusters extracted from this run has short cooling
times and would not be present in real systems.  This simulation vastly
overestimates the X-ray luminosity of the clusters.
\item[\bf Radiative:] This run includes radiative cooling using the
cooling tables of \citet{SD93} and assumes a time-varying metallicity
$Z=0.3\,(t/t_0)\Zsun$, where $t/t_0$ is the age of the Universe in
units of the current time.  Cooled material is permitted to form
stars, removing low-entropy material with short cooling times from the
centres of the clusters.  As shown in \citet{PTC00}, the inflowing gas
which replaces it has a higher net entropy and hence a higher
temperature.  A drawback of this run is that the cooling is limited
only by numerical resolution and is therefore
very ad-hoc.  In this run up to half the baryonic mass in clusters has
cooled, much more than is observed.
\item[\bf Preheating:] This run also includes radiative cooling.
Additionally, at $z=4$ the energy of each gas particle was increased
by an amount $kT=1$\,keV to crudely model energy injection at high
redshift.  This has the effect of strongly suppressing cooling so that
by the end of the simulation under 1 per cent of the gas in clusters
has cooled, much less than is observed.
\end{description}
The Radiative and Preheating runs both reproduce the observed
luminosity--temperature relation whilst having very different amounts
of cooled gas.  One might hope, therefore, that their thermal
properties also bracket those of real clusters.  We justify this
statement further in Section~\ref{sec:ent} below.

\section{Results}

\subsection{Hydrostatic equilibrium}
\label{sec:hydro}

Figure~\ref{fig:hydro} compares measures of enclosed mass versus
radius for the third-largest cluster in the Preheating simulation
(similar results are obtained for the Non-radiative and Radiative
simulations).  This particular cluster was chosen because the two
largest clusters both show signs of disturbance within $r_{500}$.  The
solid line shows enclosed mass as a function of radius.  The dashed
line shows the mass estimated from equation~(\ref{eq:masshydro}) using
the thermal pressure; a similar result is obtained using the total
(thermal plus kinetic) pressure but we have omitted this from the
plot for clarity.  

The estimated mass jiggles up and down because of variations in the
local pressure gradient.  This scatter could be much reduced by the
fitting of a smooth curve to the pressure profile, but for the
purposes of this paper we merely wish to make the point that the
clusters are in approximate hydrostatic equilibrium within \rdel.
This holds true also for the largest two clusters and for all others
that we have tested.  Note, however, that many of these other clusters
show significant departures from equilibrium within $r_{200}$.

We conclude that the use of the Equation of Hydrostatic Support to
measure masses from X-ray observations of the intracluster medium
within \rdel\ is likely to be accurate to within about 10 per cent
with no systematic offset to high or low masses.

\subsection{Temperature profiles}

\begin{figure}
\psfig{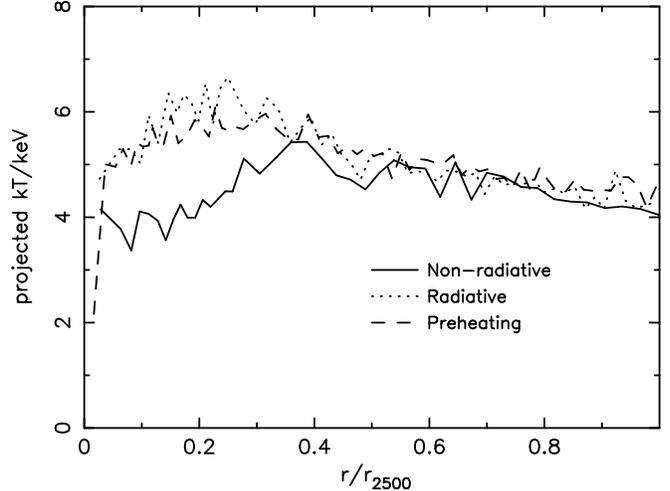}
\caption{Temperature profile of the gas within \rdel\ for one example
cluster in the three simulations.}
\label{fig:tgasprof}
\end{figure}

Figure~\ref{fig:tgasprof} shows the projected temperature profiles
within \rdel\ of the third-largest cluster in each of the three
simulations.  Note how the inflow of higher entropy material has
raised the temperature of the gas within \rdel\ in both the Radiative
and Preheating simulations relative to that in the Non-radiative
simulation, an effect first noted by \citet{PTC00}.  The temperature
profiles in this Figure differ from those shown in Figure~1 of ASF01
in that they decline slightly beyond 0.25\,\rdel\ rather than
remaining isothermal.  There are other clusters in our sample,
however, for which the temperature profiles are very similar to those
seen in the observations.

\subsection{Entropy profiles}
\label{sec:ent}

\begin{figure}
\psfig{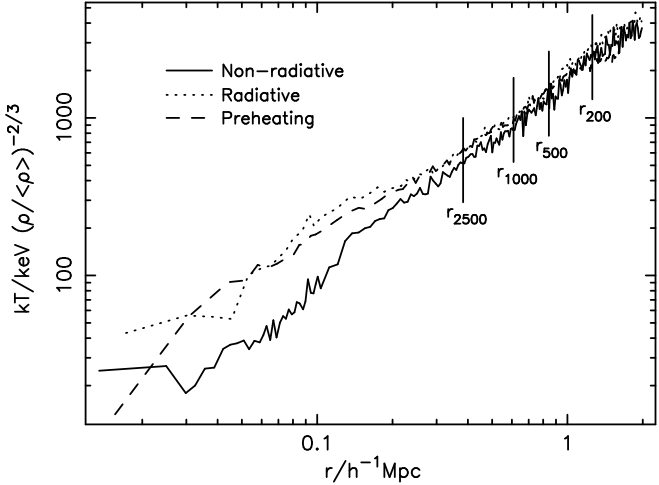}
\caption{Entropy profile of the gas for one example
cluster in each of the three simulations.}
\label{fig:entprof}
\end{figure}
The differences between the three simulations are best described in
terms of their entropy profiles.  As Figure~\ref{fig:entprof} shows,
the entropy within $r_{1000}$ in the third-largest cluster is
raised in the Radiative and Preheating runs relative to that in the
Non-radiative run.  This results in a higher gas temperature and a
less-peaked density profile.

Although the Radiative and Preheating runs have very different
cooled mass fractions, it is not surprising that they have similar
entropy profiles as we have adjusted the numerical resolution and
feedback parameters so as to reproduce the observed X-ray
luminosity--temperature relation and it has been recognized for some
time that this requires an excess of entropy in cluster cores
(\citealt{EH91}; \citealt{K91}; \citealt{B97}).  Once the entropy
profile is fixed then, assuming hydrostatic equilibrium, the cluster
temperature profile is uniquely determined.  As it happens, both the
Radiative and Preheating runs give very similar entropy profiles and
hence very similar temperature profiles for this cluster (and similar
results are obtained for other clusters).  This gives us confidence
that other models that reproduce the luminosity--temperature relation,
and in particular ones that give the correct cooled gas fraction,
would have similar thermal properties to those discussed here.

\subsection{Temperature--mass relation}

\begin{figure}
\psfig{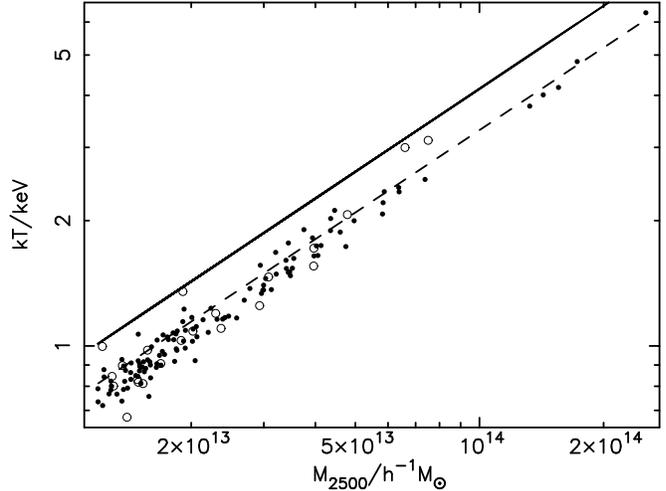}
\caption{This panel shows the mass-weighted temperature--mass relation
for gas within \rdel\ for clusters extracted from the Non-radiative
simulation.  Different symbols correspond to clusters with different
amounts of substructure, as discussed in the text.  The solid line
shows the best-fit from ASF01; the dashed line shows (extrapolated)
results from the simulations of \citet{ME01}.}
\label{fig:tgasmbad}
\end{figure}

The temperature--mass relation for the clusters extracted from the
Non-radiative simulation is shown in Figure~\ref{fig:tgasmbad}.  The
dashed line is the relation from the simulations of \citet{ME01},
extrapolated from their results for an overdensity of 500 as described
in ASF01.  Given this extrapolation our results are in good agreement
with theirs over the range for which our temperatures coincide,
$kT\approxgt1.5$\,keV.  By contrast, the relation for observed
clusters obtained in ASF01, shown by the solid line, has a
significantly lower mass normalization for a given temperature.

\begin{figure}
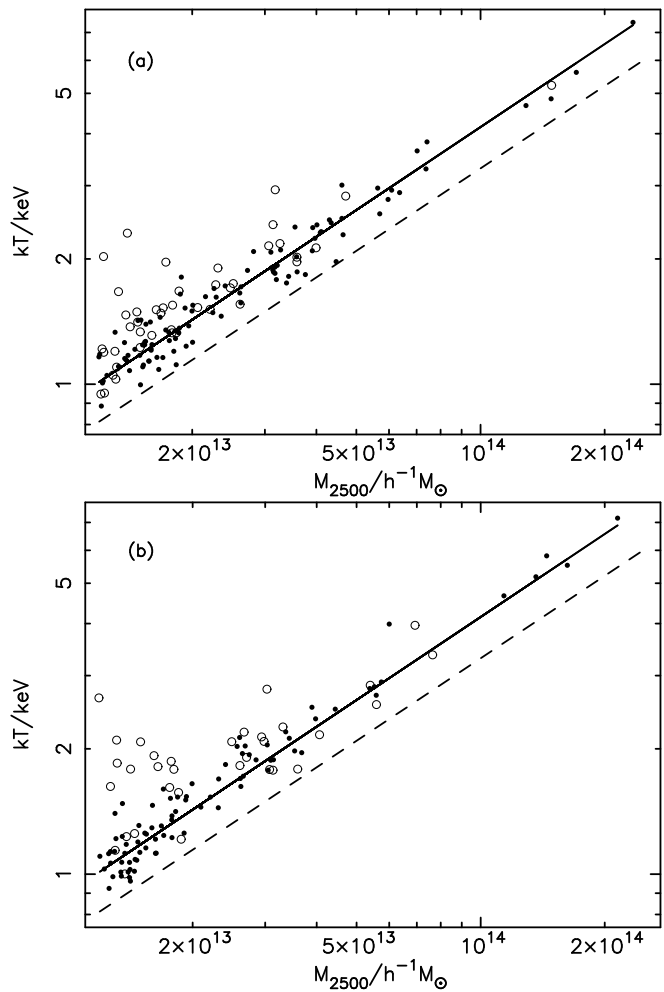

\psfig{figure=tgasm2.ps,width=8.7cm,angle=270}
\psfig{figure=tgasm3.ps,width=8.7cm,angle=270}
\caption{As for Figure~\ref{fig:tgasmbad} but for the (a) Radiative,
   (b) Preheating simulations.}
\label{fig:tgasmgood}
\end{figure}

Figure~\ref{fig:tgasmgood} shows how this relation is modified once
extra gas physics is incorporated. We see that the increase in
temperature associated with radiative cooling and/or preheating is
precisely enough to bring the simulated relation into agreement with
the observed one.

In the Figures there are several clusters that lie well above the mean
relation.  These are mostly clusters for which there is significant
velocity substructure; we indicate with open circles
those clusters for which the mean gas and dark matter velocities
within \rdel\ differ by more than 10 per cent of the rms velocity
dispersion of the dark matter.

\section{Discussion}

We have shown that simulations are capable of reproducing the observed
relationship between mass and temperature in the inner regions of galaxy 
clusters.  In particular, the mass-weighted temperature versus mass within a 
radius enclosing an overdensity of 2500 in our Radiative and Preheating
simulations agrees with the observed relation of ASF01.

There are a number of caveats, however. The temperature range of the
simulations and the observations barely overlap; we have one cluster
above 6\,keV, while ASF01 have only one below this temperature.
Nevertheless, there is no reason to suppose that our results will not
extend up to higher temperatures, though confirmation of this will
have to await resimulations of clusters drawn from larger simulation
boxes. 

Perhaps more pertinently, none of our simulations presents a
fully realistic model of clusters, the Radiative model producing too
much cooled gas and the Preheating model too little.  However they
both match the observed X-ray luminosity--temperature relation,
because they both have a higher entropy within \rdel\ than does the
Non-radiative simulation.  This increase in entropy manifests itself
as an increase in the temperature of the gas in the inner parts of the
clusters.  One might expect, therefore, that realistic clusters that
share the same entropy profile would predict the same
temperature--mass relation.

Unfortunately, the results presented in this paper and in ASF01 are of
limited use to theorists who wish to predict the temperature function
of clusters in order to constrain cosmology.  This is because they
need to relate the mass within the virial radius to the
emission-weighted temperature of clusters.  The prediction of masses
at $r_{500}$ or larger radii from the X-ray observations is a harder
problem than discussed here and will be investigated in a longer
paper.

\section*{Acknowledgements}
The simulations used in this paper were carried out on the Cray-T3E
at the EPCC as part of the Virgo Consortium programme of
investigations into the formation of structure in the Universe.  PAT
is a PPARC Lecturer Fellow; OM is supported by a DPST scholarship from
the Thai government, STK by PPARC, and ARL in part by the Leverhulme Trust.

\end{document}